\documentclass[journal,transmag]{IEEEtran}
\usepackage{mathtools}
\usepackage[pdftex]{graphicx} 
\usepackage{microtype}
\usepackage{amsmath,amssymb}
\usepackage[table,pdftex]{xcolor}
\usepackage[caption=false,font=footnotesize]{subfig} 
\usepackage[british]{babel}
\usepackage{array}
\usepackage{multirow}
\usepackage{fixltx2e} 
\usepackage{todonotes}
\usepackage{algorithm}
\usepackage{algpseudocode}
\usepackage{hyperref}

\usepackage{booktabs}
\usepackage{threeparttable}
\usepackage{footnote}
\usepackage{bbm}

\title{A new nonlocal forward model for diffuse optical tomography}

\author{Wenqi Lu$^{\dagger}$, Jinming Duan, Joshua Deepak Veesa, Iain B. Styles$^*$\\
School of Computer Science, University of Birmingham, 
Edgbaston, Birmingham, B15 2TT, UK\\
$^{\dagger}$w.lu.2@bham.ac.uk, $^*$I.B.Styles@bham.ac.uk}

\begin{document}

\maketitle

\begin{abstract}
The forward model in diffuse optical tomography (DOT) describes how light propagates through a turbid medium. It is often approximated by a diffusion equation (DE) that is numerically discretized by the classical finite element method (FEM). We propose a nonlocal diffusion equation (NDE) as a new forward model for DOT, the discretization of which is carried out with an efficient graph-based numerical method (GNM). To quantitatively evaluate the new forward model, we first conduct experiments on a homogeneous slab, where the numerical accuracy of both NDE and DE is compared against the existing analytical solution. We further evaluate NDE by comparing its image reconstruction performance (inverse problem) to that of DE. Our experiments show that NDE is quantitatively comparable to DE and is up to 64\% faster due to the efficient graph-based representation that can be implemented identically for geometries in different dimensions. \\

\end{abstract}

\section{Introduction}

\IEEEPARstart{I}n diffuse optical tomography (DOT), near-infrared light (650-900 nm) is injected into an object through optical fibers placed on its surface. The light is injected through each fibre in turn and propagates through the object. The spatial distribution of light remitted from the object's surface is measured for each source fibre, and this information is used to estimate the object's internal optical properties by iteratively refining the optical properties of a forward model of light propagation in the object until the model predictions match the measured surface remittance. As such, the forward model of light propagation must be able to accurately model the main interactions (i.e. absorption and scattering) between light and the object so as to recover internal properties faithfully. 

Technically, such interactions can be accurately described by a diffusion equation (DE) which is derived from the radiative transfer equation (RTE) \cite{duderstadt1979transport} under the assumption that the radiance in an optical medium is almost isotropic, and that the scattering interactions dominate over absorption \cite{boas1996diffuse}. Defining a computational domain $\Omega$ with boundary surface $\Gamma$ and internal domain $\Omega'$ (i.e. $\Omega {\rm{ = }}{\Omega'} \cup \Gamma $ and $\Omega ' \cap \Gamma  = \emptyset $), the DE for a continuous wave (CW) imaging system is given as
\begin{equation} \label{eq:pde}
- \nabla \cdot\left( {\kappa \left( x \right)\nabla \Phi \left( x \right)} \right) + {\mu _a}\left( {x} \right)\Phi \left( x \right) = q_0 \left(x\right)\;\;\; \rm{for}\;x \in \Omega.
\end{equation}
$\Phi\left ( x \right )$ is the photon fluence rate as a function of position $x$. The diffusion coefficient $\kappa\left( {x} \right)=1/(3(\mu_a\left( {x} \right)+\mu_s^{\prime}\left( {x} \right)))$, where $\mu_a$ and $\mu_s$ are the spatially varying absorption and scattering coefficients and $\mu_s^{\prime}\left( {x} \right)=(1-g)\mu_s\left( {x} \right)$ where $g$ is the anisotropy factor \cite{dehghani2009near}. $q_0\left ( x \right )$ is the isotropic component of the source. $\nabla$ is the gradient operator and $\nabla \cdot (\cdot)$ denotes the differential divergence of a vector function (i.e. $\kappa \nabla \Phi$). This is usually solved under the Robin boundary condition (RBC) in which light that escapes the medium does not come back. The RBC is written as 
\begin{equation} \label{eq:boundarycondition}
2A\widehat {\rm{n}}\cdot\left( {\kappa \left( x \right)\nabla \Phi \left( x \right)} \right) + \Phi \left( x \right) = 0 \;\; \rm{for}\;x \in \Gamma,
\end{equation}
where $\widehat {\rm{n}}$ denotes the outward unit normal on the boundary. $A$ is related to the relative refractive index mismatch between the medium and air and is derived from Fresnel's law \cite{dehghani2009near}.

Mathematically, Equation (\ref{eq:pde}) is an elliptic partial differential equations, the differential operators (i.e. gradient or divergence) in which are defined using the classical vector calculus. A general approach to analytically solve the DE (with its RBC) is to apply the Green function, but analytical solutions are only known for homogeneous objects \cite{farrell1992diffusion,arridge1999optical,allen1991modified}. For more complex DOT geometries, the finite element method (FEM) \cite{arridge1999optical} is commonly used to discretize the DE and its RBC. In this discretization, the computational domain $\Omega$ is divided into a series of elements (triangles in 2D, tetrahedra in 3D)\footnote{One normally terms the discretized geometry using FEM as a FE mesh}. However, FEM implementations can be difficult and time-consuming, especially when higher-order polynomial basis (shape) functions are used for non-linear interpolation between vertices of high-order elements \cite{lynch2004numerical}. In previous work \cite{lu2019graph}, we introduced a graph representation to discretize a total-variation regularization term for the inverse problem in DOT. In this discretization, the object geometry is represented by an unstructured graph, defined by vertices, edges and weights. The graph was constructed by exploring neighborhood relationships between vertices.

In order to fully leverage the power of graph-based discretization, one must use the nonlocal vector calculus. In the classical local vector calculus, the differential operators are numerically evaluated using purely local information. In the nonlocal calculus, the operators include more pixel information in the domain. For example, in image processing, some well-known PDEs and variational techniques such as nonlocal image denoising \cite{buades2005non,gilboa2008nonlocal}, segmentation \cite{bresson2008non} and inpainting \cite{duan2013color,duan2015fast} have explored the advantages of nonlocal vector calculus \cite{gilboa2008nonlocal,gunzburger2010nonlocal}. When applied to these problems, local operators include information from only neighbouring pixels whilst nonlocal methods include information from a wider area and are naturally formulated in a graph-based representation instead of in terms of the classical local differential operators.

In image processing, nonlocal methods are shown to have several advantages over local methods, including preservation of important image features such as texture and ability to handle unstructured geometries. It has also been observed that many PDE-based physical processes, minimizations and computational methods, such as CT image processing and reconstruction \cite{zhang2017applications,zeng2015spectral}, can be generalized to be nonlocal. Therefore we expect that such a framework may be useful for the physical modelling in DOT.

As such, we propose a nonlocal diffusion equation (NDE) as a new forward model for DOT. The concept of differential operators under the nonlocal vector calculus \cite{gilboa2008nonlocal,gunzburger2010nonlocal,duan2013color,duan2015fast} is used to formulate a new forward model that can accurately simulate light propagation in turbid media. The discretization for the NDE is performed using a graph-based numerical method (GNM). As a result, the proposed method naturally applies without modification to complex, unstructured DOT geometries in both two and three dimensions. The accuracy of the proposed model is compared against the conventional diffusion equation implemented by FEM and to the existing analytical solution on a homogeneous slab. We also compare the image reconstruction accuracy of different forward models on a 2D circular model and a 3D human head model. 

\section{METHODOLOGY}
\label{sec:examples}
Our approach is based on reformulating the diffusion equation (Equation (\ref{eq:pde})) in terms of nonlocal differential operators. We denote $\nabla_w(\cdot)$, $\mathrm{div}_w(\cdot)$ and ${\cal N}_w(\cdot)$ as the nonlocal gradient, the nonlocal divergence and the nonlocal normal derivative, respectively. Their definitions are given in Equations (\ref{eq:nonlocalgradient}), (\ref{eq:nonlocaldivergence}) and (\ref{eq:NLNormalDeriv}). We simply replace the differential operators in Equation (\ref{eq:pde}) with their nonlocal counterparts and solve the new NDE under the framework of nonlocal vector calculus:
\begin{equation} \label{eq:NLpde}
- \mathrm{div}_w \left( {\kappa \left( x \right)\nabla_w \Phi \left( x \right)} \right) + {\mu _a}\left( x \right) \Phi \left( x \right) = q_0 \left(x\right)\;\; \rm{for}\;x \in \Omega '.
\end{equation}
Similarly, we reformulate the RBC with the nonlocal normal derivative and the nonlocal gradient to give a nonlocal boundary condition (NBC):
\begin{equation} \label{eq:NLboundary}
2A{\cal N}_w\left( {\kappa \left( x \right)\nabla_w \Phi \left( x \right)} \right) + \Phi \left( x \right) = 0 \;\; \rm{for}\;x \in \Gamma.
\end{equation}
We now formulate a graph-based numerical method to discretize the NDE with its NBC. Following established methods \cite{gilboa2008nonlocal,gunzburger2010nonlocal}, we first discretize the computational domain $\Omega$ using a weighted graph $G = \left( {V,E,w} \right)$, where $V = \left\{ {{V_k}} \right\}_{k = 1}^N$ denotes a finite set of $N$ vertices, and $E \in V \times V$ represents a finite set of weighted edges. Here $V=V_{\Omega'} \cup V_{\Gamma}$ with $V_{\Omega'}$ representing vertices in $\Omega'$ and $V_{\Gamma}$ vertices on boundary $\Gamma$. In this study, we assume that $G$ is an undirected simple graph (no multiple edges). Let $(i,j) \in E$ be an edge of $E$ that connects the vertices $i$ and $j$ in $V$. The weight $w_{ij}$ denotes the similarity between two vertices $i$ and $j$. The computation of this quantity is discussed later in this section. The nonlocal differential operators required by Equations (\ref{eq:NLpde}) and (\ref{eq:NLboundary}) on the graph $G$ are then defined as follows.

\textbf{Definition} (\textit{Nonlocal gradient}). For a function $\Phi_i: V \to \mathbb{R}$ and a nonnegative and symmetric weight function $w_{ij}$: $V \times V \to \mathbb{R}$, the nonlocal partial derivative can be written as
\begin{equation} \label{eq:nonlocalderivative}
{\partial_j}{\Phi _{i}} \triangleq \left( {{\Phi _j} - {\Phi _i}} \right)\sqrt {{w_{ij}}} :V \times V \to \mathbb{R}.
\end{equation}
Therefore the nonlocal gradient ${\nabla_w}{\Phi _{i}}$ is defined as the vector of all partial derivatives:
\begin{equation} \label{eq:nonlocalgradient}
{\nabla_w}{\Phi _{i,j}} \triangleq \left( {{\Phi _j} - {\Phi _i}} \right)\sqrt {{w_{ij}}} {\text{       }} :V \times V \to \mathbb{R}.
\end{equation}

\textbf{Definition} (\textit{Nonlocal divergence}). Given a vector function $\boldsymbol{\nu}_i$: $V_{\Omega'} \to \mathbb{R}$ and a weight function $w_{ij}$: $V \times V \to \mathbb{R}$, the nonlocal divergence operator $\mathrm{div}_w$ acting on $\boldsymbol{\nu}_i$ is  
\begin{equation} \label{eq:nonlocaldivergence}
\mathrm{div}_w\, \boldsymbol{\nu} _i \triangleq \sum\limits_{j=1}^N {\left( {{\nu _{ij}} - {\nu _{ji}}} \right)\sqrt {{w_{ij}}} } {\text{ }}: V_{\Omega'} \to \mathbb{R}
,\end{equation}
where ${\nu _{ij}}$ is the $j$'th element of $\boldsymbol{\nu} _{i}$. 

\textbf{Definition} (\textit{Nonlocal normal derivative}). Given a function $\boldsymbol{\nu}_i$: $V_{\Gamma} \to \mathbb{R}$ and a weight function $w_{ij}$: $V \times V \to \mathbb{R}$, the nonlocal normal operator acting on $\boldsymbol{\nu}_i$ is
\begin{equation} \label{eq:NLNormalDeriv}
{\cal N}_w \boldsymbol{\nu}_i \triangleq - \sum\limits_{j=1}^N{\left( {\nu _{ij} - {\nu _{ji}}} \right)\sqrt{w_{ij}}} {\text{ }}: V_{\Gamma} \to \mathbb{R} .
\end{equation}

\textbf{Definition} (\textit{Nonlocal Laplacian}). Let $\Phi_i: V \to \mathbb{R}$ and $w_{ij}$: $V \times V \to \mathbb{R}$. The linear nonlocal Laplace operator acting on $\Phi_i$ is defined based on Equation (\ref{eq:nonlocalgradient}) and (\ref{eq:nonlocaldivergence}): 
\begin{equation} \label{eq:nonlocallap}
{\Delta_w}{\Phi _{i}} \triangleq \frac{1}{2}\mathrm{div}_w\,\left( {{\nabla}_w{\Phi _{i}}} \right) = \sum\limits_{j =1}^N {\left( {{\Phi _{j}} - {\Phi _{i}}} \right){w_{ij}}} {\text{ }}: V \to \mathbb{R}.
\end{equation}
The nonlocal normal derivative ${\cal N}_w$ in Equation (\ref{eq:NLNormalDeriv}) is a nonlocal analogue of the normal derivative operator at the boundary encountered in the classical differential vector calculus (i.e. $\widehat {\rm{n}}$ in Equation (\ref{eq:boundarycondition})). Note that $\mathrm{div}_w$ in Equation (\ref{eq:nonlocaldivergence}) and ${\cal N}_w$ in Equation (\ref{eq:NLNormalDeriv}) have similar definitions but differ in their signs and the regions over which $\mathrm{div}_w\, \boldsymbol{\nu}_i$ and ${\cal N}_w \boldsymbol{\nu}_i$ are calculated. Also note that the mapping $\boldsymbol{\nu}_i \mapsto {\cal N}_w \boldsymbol{\nu}_i$ is scalar-valued which is analogous to the local differential divergence of a vector function in Equation (\ref{eq:pde}). Finally, with the definitions of $\mathrm{div}_w$ and ${\cal N}_w$, the nonlocal divergence theorem is $\int_{\Omega '} {\mathrm{div}_w\, \boldsymbol{\nu}dx}  = \int_\Gamma  {{\cal N}_w \boldsymbol{\nu}dx}$, which essentially relates the flow (i.e. flux) of a nonlocal vector field through a boundary/surface to the behaviour of the nonlocal vector field inside the boundary/surface.

It should be noticed from the nonlocal differential operator definitions (Equations \ref{eq:nonlocalgradient}, (\ref{eq:nonlocaldivergence}), (\ref{eq:NLNormalDeriv}) and (\ref{eq:nonlocallap})) that, in a full non-local scheme, each vertex has connections with all the vertices in $V$ over $\Omega$ such that the constructed graph is fully connected. This can make the computational load extremely heavy and so approaches based on spectral graph theory \cite{bertozzi2012diffuse,merkurjev2013mbo} or nearest neighbors \cite{bresson2014multi}, are typically employed to partition the vertices in the computational domain into groups according to their similarities. For example, Bertozzi \cite{merkurjev2013mbo} used spectral approaches along with the Nystr\"om extension method to efficiently calculate the eigendecomposition of a dense graph Laplacian. The second eigenvector of the graph Laplacian was used to initialize the partitioning so that the weights between vertices in different groups are small and the weights between vertices within the same group are large. In this paper, we build the graph by using the positions of the nodes and the connectivity between nodes in the finite element mesh as the vertices and edges in the graph to sparsify the graph for computational efficiency. We have learned from previous work \cite{lu2019graph} that the graph-based nonlocal inverse model with this sparse method can achieve accurate and stable reconstruction, regardless of the mesh resolution. Therefore for each vertex $i$, we consider only those vertices that are directly connected to the vertex $i$ for ${{\cal N}_i}$ (i.e. those vertices that share the same edge with $i$). With this structure and the nonlocal discrete differential operators, we can derive the following discretized versions of Equations (\ref{eq:NLpde}) and (\ref{eq:NLboundary}):
\begin{equation} \label{eq:discretisation}
\begin{split}
\sum\limits_{j \in {{ {\cal{N}}}_i}} {\left( {{\kappa _i} + {\kappa _j}} \right)\left( {{\Phi _i} - {\Phi _j}} \right){w_{ij}}}  + {{\mu_a}_i}{\Phi _i} = {q_{0i}}\;\;\;{\text{for}}\;i \in \Omega ' \hfill \\
2A\sum\limits_{j \in {{{\cal{N}}}_i}} {\left( {{\kappa _i} + {\kappa _j}} \right)\left( {{\Phi _i} - {\Phi _j}} \right){w_{ij}}}  + {\Phi _i} = 0\;\;\;\;{\text{for}}\;i \in \Gamma  \hfill \\ 
\end{split}
\end{equation}

The nonnegative and symmetric weight function $w_{ij}$ between two connected vertices $i$ and $j$ has many possible choices. In this work, we first obtain the similarity $w_{ij}$ by simply using the inverse of the Euclidean distance $d_{ij}$ between two nodes. Then we normalize the similarity using $w_{ij}/{\sum_{j \in {{{\cal{N}}}_i}}w_{ij}}$ to convert the similarities into probabilities and ensure that the probabilities sum to one. 


We note that due to the nature of the graph representation, the implementation of Equation (\ref{eq:discretisation}) is identical for a 2D or 3D geometry. It should also be noted that increasing the number of vertices and edges will decrease the sparsity of the graph and increase the computational burden with no change in the implementation. Under these assumptions, Equation (\ref{eq:discretisation}) can be rewritten in matrix form as
\begin{equation} \label{eq:NLLinearSystem}
{\cal{M}}\Phi  = Q.
\end{equation}
${\cal{M}}$ is a $N \times N$ sparse matrix and a symmetric, diagonally dominant and positive definite real-value matrix, whose entries are
\[{{\cal{M}}_{i,j}} = \left\{ \begin{array}{ll}
\sum\limits_{j \in {{\cal {N}}_i}}^{} {\left( {{\kappa _i} + {\kappa _j}} \right){w_{ij}}}  + {{\mu_a}_i}  & {\rm{if}}\; i=j \in \Omega '\\
\sum\limits_{j \in {{\cal {N}}_i}}^{} {\left( {{\kappa _i} + {\kappa _j}} \right){w_{ij}}}  + \frac{1}{2A}  & {\rm{if}}\; i=j \in \Gamma \\
-{\left( {{\kappa _i} + {\kappa _j}} \right){w_{ij}}}  & {\rm{if}}\; i \neq j \; {\rm{and}}\; {j \in {{\cal {N}}_i}} \\
 \;\;\;\;\;\;\;\;\;\;\;{0} & {\rm{otherwise}}
\end{array} \right..\]
$Q$ is a $N \times N_s$ sparse matrix where $N_s$ is the number of sources and each column represents one distributed Gaussian source. The linear system (Equation (\ref{eq:NLLinearSystem})) can be solved exactly by using a direct solver with Cholesky decomposition.

\section{EXPERIMENTAL RESULTS}
In this section, numerical experiments are conducted to quantitatively evaluate the performance of the proposed NDE method. The NDE method with the GNM implementation will be compared against the original DE with the FEM implementation. We evaluate the light propagation performance of the proposed method in a 3D homogeneous rectangular-slab where the analytical solution is known, followed by two dimensional (2D) and three dimensional (3D) image reconstruction examples. All the experiments are performed using Matlab 2018b on a Windows 7 platform with an Intel Xeon CPU i7-6700 (3.40 GHz) and 64 GB memory. 

\subsection{Forward Modelling on A 3D Homogeneous Rectangular-slab Model}
\begin{figure}[htbp]
\centering
\includegraphics[width=0.8\linewidth]{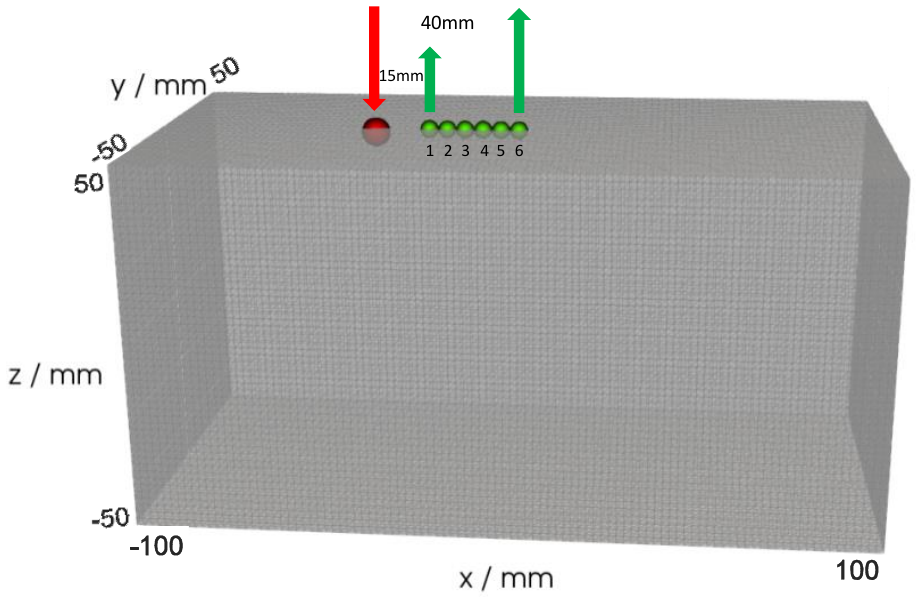}
\caption{Rectangular-slab mesh with one source (red dot) and six detectors (green dots). The distance between the source and the six detectors varies from 15 mm to 40 mm, in 5 mm increments.}
\label{fig:f1}
\end{figure}

To quantitatively compare our GNM method with classical FEM approaches, we model a homogeneous rectangular-slab of size: 200$\times$100$\times$100 mm$^3$, as shown in Figure~\ref{fig:f1}. The mesh is composed of 442381 nodes corresponding to 2620541 tetrahedral elements, with the average nodal distance of 1.5 mm. For the forward model based on FEM, such a discrete structure can be directly employed for the finite element method. However, the forward model based on GNM requires only the vertices and edges of the mesh. The optical parameters $\mu_a$ and $\mu_s^{\prime}$ in this slab were set to 0.01 mm$^{-1}$ and 1 mm$^{-1}$, respectively. We conduct simulations using a CW source for which we can analytically calculate the photon flux measurement on the boundary (BF) as well as the fluence rate (FR) at each vertex. The analytical solutions are then compared with the solutions from the forward model based on FEM and GNM. 

The analytical solution of the BF has the form \cite{farrell1992diffusion}:
\begin{equation} \label{eq:analyticalintensity}
\begin{split}
 I\left( \rho  \right) = \frac{1}{{4\pi }}\left[ {\frac{1}{{{\mu _a} + {\mu _{s}^\prime}}}}\left( {{\mu _\mathrm{eff}} + \frac{1}{{{r_1}}}} \right)\frac{{{e^{ - {\mu _\mathrm{eff}}{r_1}}}}}{{r_1}^2} \right. \\
\left. + \frac{{3 + 4A}}{{3\left( {{\mu _a} + {\mu _{s}^\prime}} \right)}}\left( {{\mu _\mathrm{eff}} + \frac{1}{{{r_2}}}} \right)\frac{{{e^{ - {\mu _\mathrm{eff}}{r_2}}}}}{{r_2}^2} \right] ,
\end{split}
\end{equation}
where $\rho$ represents the distance from the source, $A$ is the internal reflection parameter for the air-tissue interface, ${\mu _\mathrm{eff}}$ is the effective attenuation coefficient which is $\sqrt {3{\mu _a}\left( {{\mu _a} + {\mu _{s}^\prime}} \right)} $, ${r_1} = \sqrt {1/( {\mu _a} + {\mu _{s}^\prime} )^2 + {\rho ^2}} $ and ${r_2} = \sqrt {{( 3 + 4A )^2}/{( {3( {\mu _a} + {\mu _{s}^\prime} )} )^2} + {\rho ^2}} $. 

\begin{figure}[htbp]
\centering
\includegraphics[width=1\linewidth]{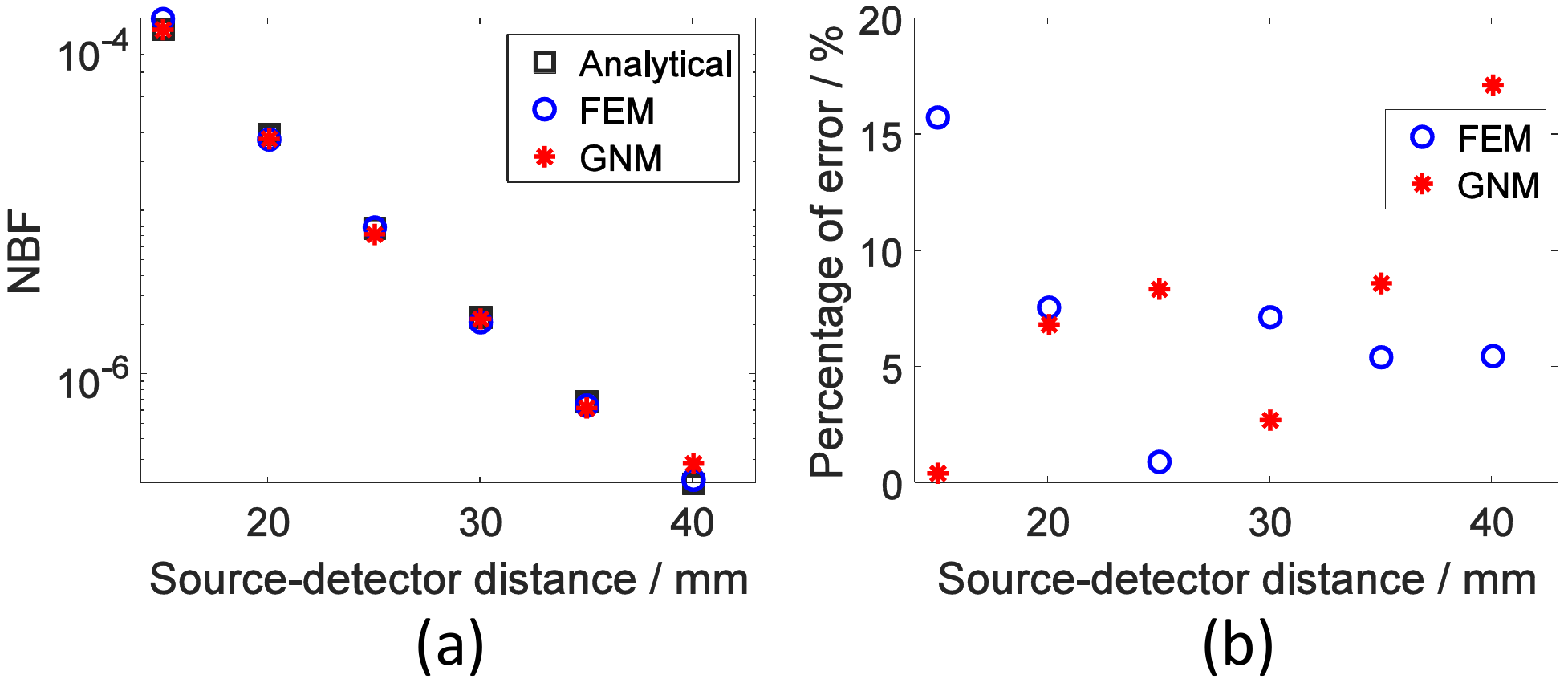}
\caption{The flux measurements on the boundary versus the source-detector distance. (a): NBF; (b): Percentage of error based on NBF.}
\label{fig:f2}
\end{figure}

In Figure~\ref{fig:f2} (a), we plot the normalized photon flux at the boundary (NBF). We normalize the BF to remove any constant offset resulting from the use of different propagation models. It can be seen that the NBF from both forward models match the analytical solution. In order to observe the difference clearly, in Figure~\ref{fig:f2} (b), we plot the percentage of error between the analytical solution and the other two methods with regards to NBF. The percentage of error is calculated by, for each source-detector channel, dividing the absolute difference between each forward model and the analytical solution by the analytical solution. We average the percentage errors along the six source-detector pairs. The forward models based on FEM and GNM are both shown to reproduce the analytical solution to within 7\% on average. 

We then compare the FR calculated at the vertices inside of the medium. The analytical solution of the FR is \cite{allen1991modified}:
\begin{equation} \label{eq:analyticalfluencerate}
\begin{split}
\Phi \left( {r,z} \right) = \frac{P\mu_\mathrm{eff} ^2}{4\pi\mu_a}\left[ \left( {\frac{\exp \left\{ { - \mu_\mathrm{eff} {{\left[ {{{\left( z - z_0 \right)}^2} + r^2} \right]}^{1/2}}} \right\}}{{ - {\mu _\mathrm{eff}} {{\left[ \left( z - z_0 \right)^2 + r^2 \right]}^{1/2}}}}} \right) \right.\\
\left. - \left( {\frac{{\exp \left\{ { - \mu_\mathrm{eff} {{\left[ \left( z + z_0 \right)^2 + {r^2} \right]}^{1/2}}} \right\}}}{{ - \mu_\mathrm{eff} {{\left[ {{{\left( z + z_0 \right)}^2} + r^2} \right]}^{1/2}}}}} \right) \right],
\end{split}
\end{equation}
where $P$ is the source power. $z_0$ is the depth of the source which is $1/\mu'_s$.  $z$ represents the depth under the surface which is $z=50$mm in our case. $r$ is the distance between a given vertex and the source on the X-Y plane. Note that $\sqrt {{( z - {z_0} )^2} + {r^2}} $ represents the distance between a given vertex and the source. 

\begin{figure}[htbp]
\centering
\includegraphics[width=1\linewidth]{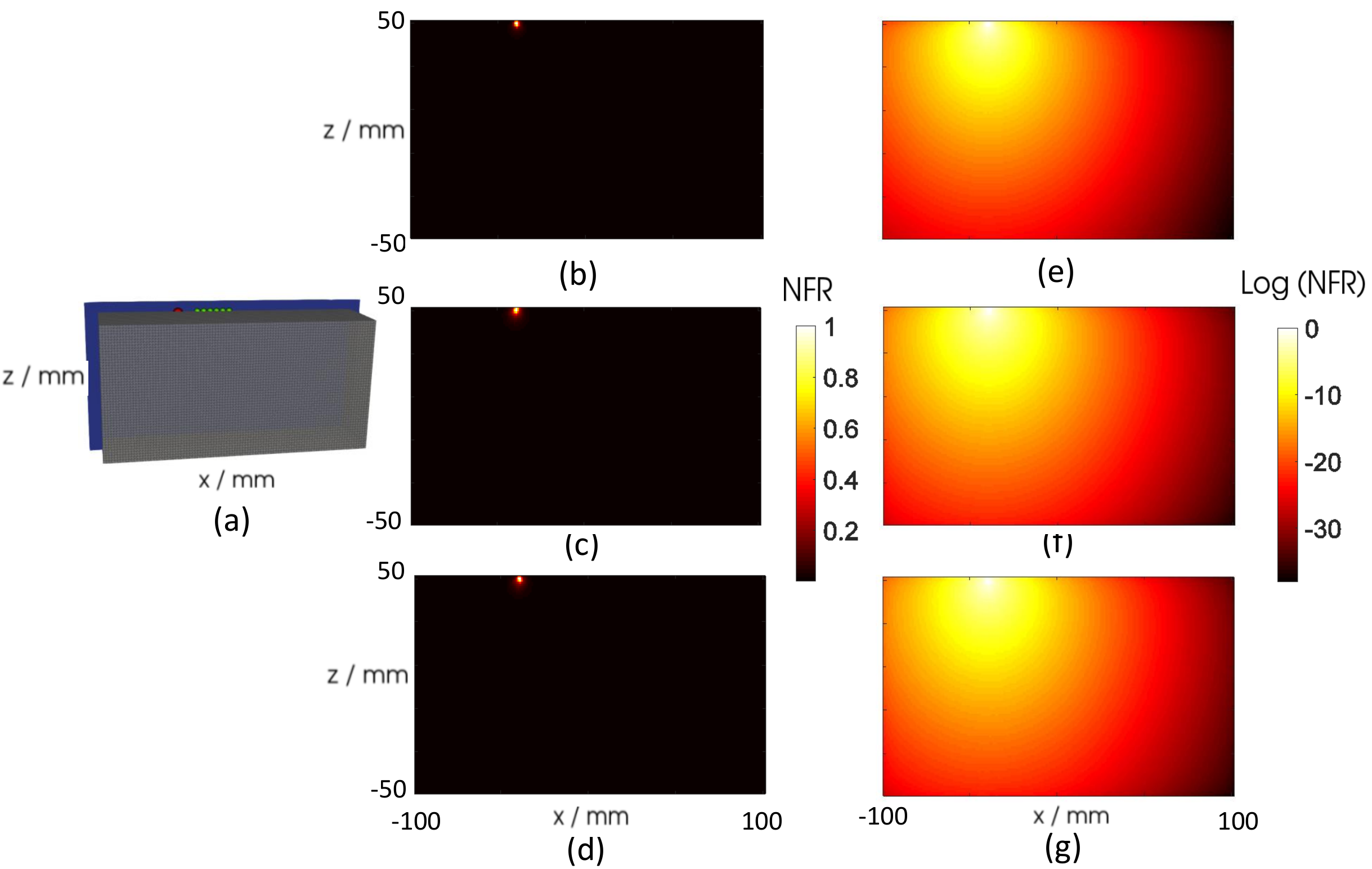}
\caption{(a) The rectangular-slab mesh with the ROI (blue triangular); (b)-(d): NFR at each vertex in the ROI calculated using the analytical solution, forward models based on FEM and the one based on GNM, respectively; (e)-(g): logarithm of the NFRs, corresponding to (b)-(d).}
\label{fig:f3}
\end{figure}

\begin{figure}[htbp]
\centering
\includegraphics[width=0.8\linewidth]{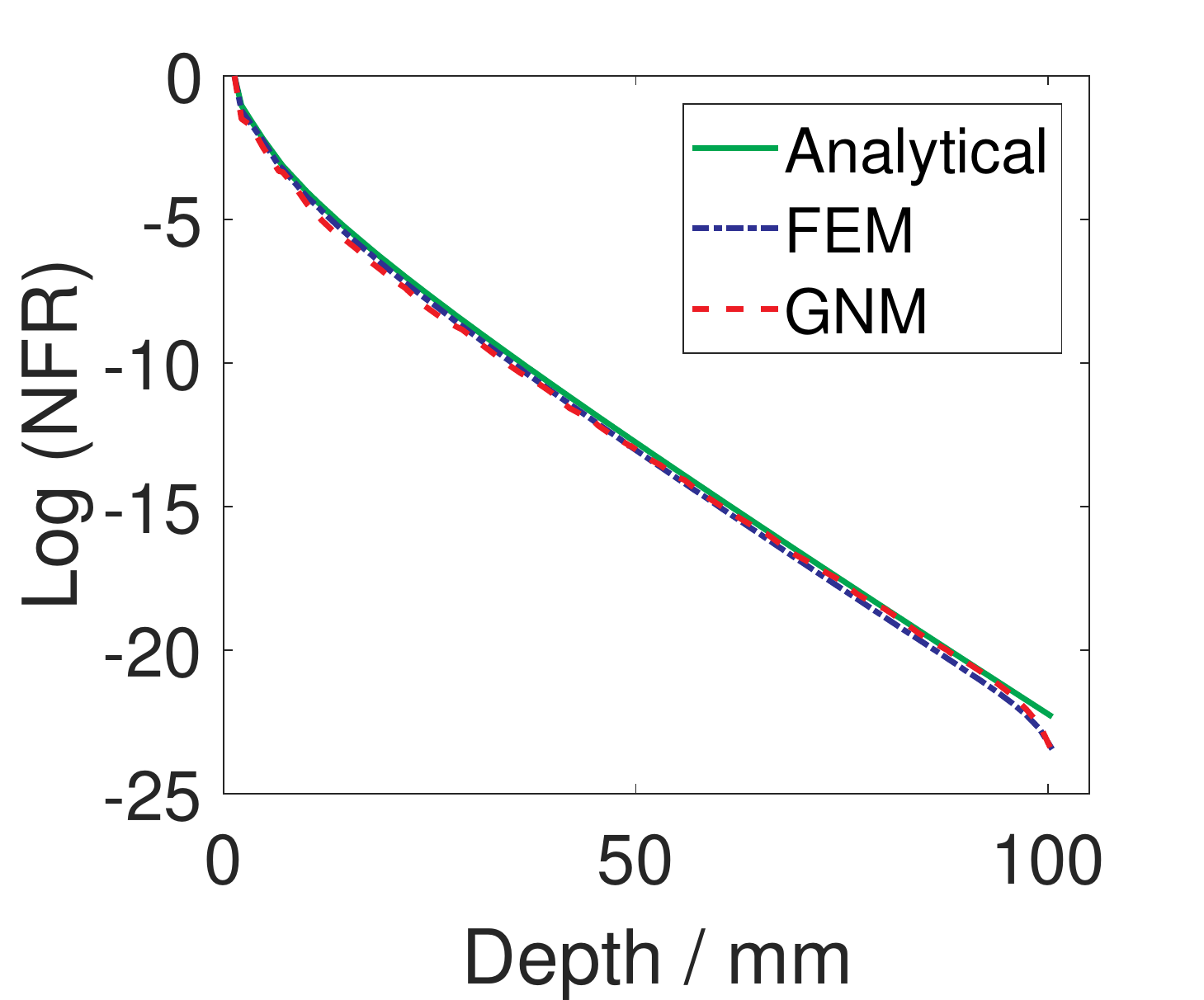}
\caption{Descending tendency of the NFR from the source to the medium along the z axis.}
\label{fig:f7}
\end{figure}

In Figure~\ref{fig:f3}, we compare the FR calculated using Equation (\ref{eq:analyticalfluencerate}) and the FEM and GNM methods. The blue rectangular area in Figure~\ref{fig:f3}(a) represents the position of the region of interest (ROI). This area is positioned across the sources and detectors, and parallel to the X-Z plane. For each method, in order to remove any constant offset resulting from the use of different propagation models, we rescaled FR onto the range $[0,1]$ by dividing the FR with the highest FR value in the ROI and name the rescaled FR as NFR. This is necessary because in FEM, point sources are distributed across the nodes belonging to the element in which the source is placed, whereas in GNM, the source is fully attached to the nearest vertex. The two methods can therefore have different initialization states for the same source. In Figure~\ref{fig:f3}(b)-(d), we plot the NFR at each vertex in the ROI calculated using the analytical method, and the FEM and GNM models, respectively. We also plot its logarithm in (e)-(g), corresponding to  the NFR in (b)-(d) respectively. It can be observed that the light propagation in the medium modelled by the proposed forward model is comparable to the one modelled by the forward model based on FEM. In order to see the difference clearly, in Figure~\ref{fig:f7}, we plot the descending tendency of the NFR calculated by different propagation methods. Specifically, we plot the logarithm of NFR along the z axis starting from the source position. As can be seen, for all methods the fluence rate gradually drops as the light penetrates deeper. The descending tendency of the curves derived from the both forward methods are almost parallel to the one from the analytical solution. Therefore we can see that all the three models can generate the same NFR distribution.

\begin{figure}[htbp]
\centering
\includegraphics[width=1\linewidth]{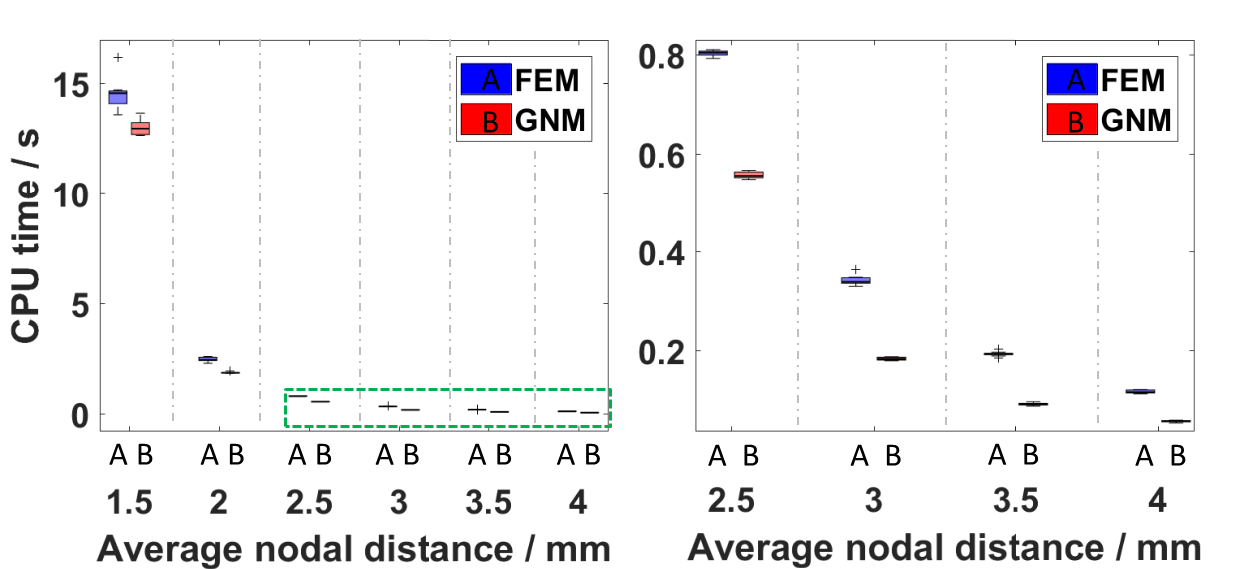}
\caption{CPU time (s) consumed at one source-detector channel using different forward models. 'A' represents the FEM approach while 'B' represents the GNM approach. Right figure is the zoomed-in plot of the area in the green dash line of the left figure.}
\label{fig:f8}
\end{figure}

After evaluating the accuracy of the fluence rates and boundary measurements modelled by different forward models, in Figure~\ref{fig:f8}, we compare the computational efficiency of FEM and GNM  forward models. We run each model on six meshes with different average nodal distance of 1.5, 2, 2.5, 3, 3.5 and 4 mm respectively. The mesh spatial resolution becomes lower when the nodal distance is larger. We run each forward modelling process ten times and record the mean and standard deviation of the CPU time consumed for computing one source-detector channel. For a fair comparison, we use a direct solver with Cholesky decomposition to solve the linear equation resulting from each forward model. For all mesh resolutions, based on each source-detector channel, the CPU times required by the FEM model are larger than that by GNM. When the mesh resolution is low (for example the case where the average nodal distance is 4 mm) the CPU time consumed by the FEM approach (0.11s) is 175\% larger than the time required by the GNM approach (0.04s). 
When the mesh resolution is high (average nodal distance is 1.5 mm), the CPU time consumed by the FEM approach (14.6s) is only 14\% longer than the GNM approach (12.7s). This finding demonstrates the computational efficiency of the proposed forward model.

\subsection{Image Reconstruction Using Different Forward Models}
We now consider the recovery of the optical properties at each vertex within the medium using both forward models. The image reconstruction process is implemented by iteratively refining the optical properties of the forward model until the forward model prediction matches the boundary measurements \cite{dehghani2009near}. It can be implemented by solving the following minimization problem:
\begin{equation} \label{eq:inverse11}
{\mu_a ^*} = \arg {\min _{\mu_a} }\left\{\| {\Phi _{}^{\rm{M}} - {\cal F}\left( \mu_a  \right)\|_2^2 + \lambda {\cal R}\left( \mu_a  \right)} \right\},
\end{equation}
where $\Phi^{\rm{M}}$ represents the boundary measurements acquired from the optical detectors, ${\cal F}$ is the non-linear operator induced from the forward model, ${\cal R}$ is a general regularization term, and $\lambda$ is a weight that determines the extent to which regularization will be imposed on the solution $\mu^*$. In this paper, we adopt the popular quadratic Tikhonov-type regularization (${\cal R}(\mu_a)=\|\mu_a - \mu_{a,0} \|_2^2$) for all methods for fair comparison \cite{dehghani2009near}. Four quantitative evaluation metrics are considered to evaluate the reconstruction results: the average contrast (AC) \cite{lu20181}, peak signal-to-noise ratio (PSNR) \cite{lu20181}, structural similarity index (SSIM) \cite{lu2016implementation} and root mean square error (RMSE) \cite{lu2016implementation}. If the reconstructed image is identical to the ground truth image, AC is equal to 1. For PSNR and SSIM, the recovered image has higher quality if higher PSNR or SSIM values are obtained. Lower RMSE represents better reconstruction results. Randomly generated Gaussian noise is added to the amplitude of the measurement vector to simulate real noise in a CW system. In order to reduce the randomness resulting from the randomly distributed Gaussian noise, we run each experiment ten times and record the average (mean) and standard deviation (SD) of the four evaluation metrics.

\subsubsection{Image Reconstruction on A Homogeneous Circular Model}
\begin{figure}[h]
\centering
\includegraphics[width=0.45\textwidth]{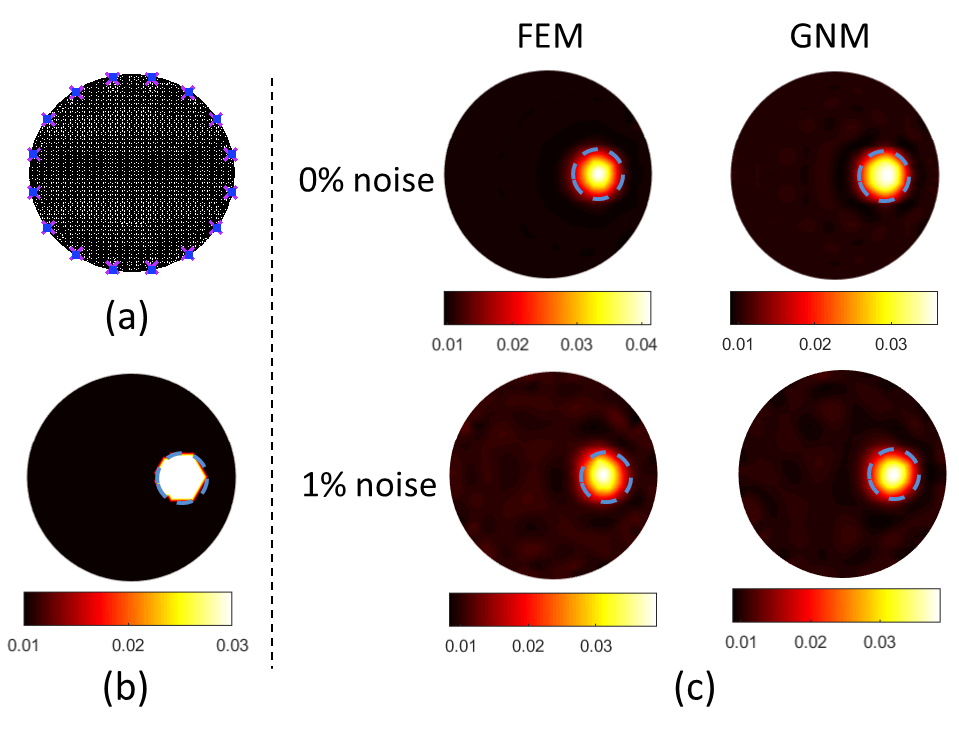}
\caption{(a): A typical circle mesh with sixteen co-located sources and detectors; (b): True distribution of $\mu_a$; (c): Images reconstruction of $\mu_a$ using the forward model based on FEM and GNM (from left to right column) on $0\%$ (top part) and $1\% $ (bottom part) noisy data.}
\label{fig:SD}
\end{figure}

We consider a 2D homogeneous circular geometry containing one target activation region (Figure~\ref{fig:SD} (b)). The  model has a radius of 43mm and is composed of 1785 nodes and 3418 linear triangle elements. Sixteen source-detector fibres are placed equidistant around the external boundary for data acquisition  (Figure~\ref{fig:SD} (a)). When one fibre as a source is turned on, the rest are used as detectors, leading to 240 total boundary measurements. All sources were positioned one scattering distance within the outer boundary because the source is assumed to be spherically isotropic. The background absorption coefficient is set to 0.01 mm$^{-1}$. One 10mm radius target region is centred at (20mm, 0mm) with 0.03 mm$^{-1}$ absorption coefficient. The reduced scattering coefficient is set to be homogeneous throughout the whole computational domain with the value of 1 mm$^{-1}$. 1\% normally distributed Gaussian noise was added to the amplitude of the measurement vector.

\begin{figure}[h]
\centering
\includegraphics[width=0.45\textwidth]{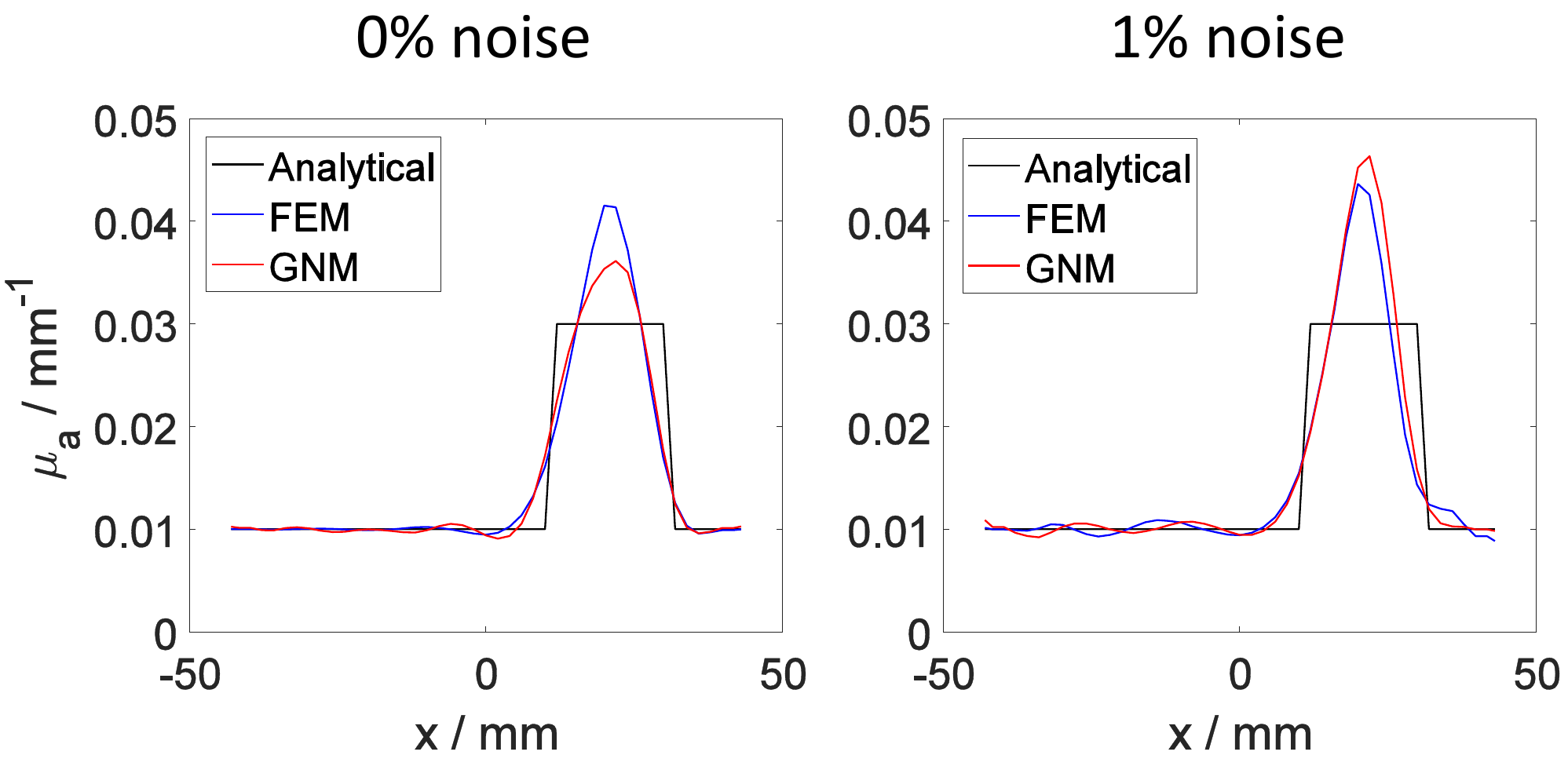}
\caption{1D cross sections of images recovered in Figure~{\ref{fig:SD}} along the horizontal line across the centre of the target. Left to right column: $0\%$ and $1\%$ added Gaussian noise.}
\label{fig:cross_section}
\end{figure}

Figure~\ref{fig:SD} (c) shows the reconstruction results using the forward model based on FEM (Equation \ref{eq:pde}) and GNM (Equation \ref{eq:NLpde}) on $0\% $ and $1\% $ noisy data respectively. By visual inspection, it is evident that for the same level of Gaussian noise, the image recovered using the GNM approach is similar to the one recovered using the FEM approach. Figure~\ref{fig:cross_section} gives the 1D cross section of the results recovered in Figure~\ref{fig:SD} along the horizontal line across the centre of the target (20mm, 0mm). It can be seen that the curves resulting from different forward models have similar edge smoothing resulting from the Tikhonov regularization and slightly different peak values. This is consistent with our visual observation from the reconstructed images in Figure~\ref{fig:SD}. 

In Table~\ref{tb:CIRCLE_table}, the values of the metrics AC, PSNR, SSIM and RMSE are shown to qualitatively evaluate the results in Figure~{\ref{fig:SD}}. It can be observed that when the data is clean,  GNM gives AC closer to 1, slightly higher PSNR and SSIM, and lower RMSE than FEM. For the noisy data, GNM achieves similar AC, PSNR, SSIM and RMSE values with the FEM approach. This experiment quantitatively validates the forward modelling capacity of our proposed model and the consistency between these two forward models.

\begin{table}[htbp]
\centering 
\caption{\bf Evaluation metrics for the recovered results using FEM and GNM on data with 0\% and 1\% added noise.}
\label{tb:CIRCLE_table}
\begin{tabular}{c|c|c|c|c}
\hline
     & \multicolumn{2}{c|}{0\% noise} & \multicolumn{2}{c}{1\% noise (Mean $\pm$ SD)}    \\ \hline
     & FEM            & GNM           & FEM             & GNM             \\ \hline
AC   & 1.1           & 1.0          & 1.1 $\pm$ 0.1       & 1.1 $\pm$ 0.1      \\ \hline
PSNR & 54.5          & 55.8         & 54.3 $\pm$ 0.5      & 54.3 $\pm$ 0.4      \\ \hline
SSIM & 99.6e-2     & 99.7e-2      & 99.6e-2 $\pm$ 7.6e-4 & 99.6e-2 $\pm$ 4.9e-4 \\ \hline
RMSE & 1.9e-3         & 1.6e-3        & 1.9e-3 $\pm$ 1.2e-4 & 1.9e-3 $\pm$ 8.4e-5 \\ \hline
\end{tabular}
\end{table}

\subsubsection{Image Reconstruction on A Heterogeneous Head Model}
We now evaluate both forward models on a physically realistic three dimensional heterogeneous head model. This head model is composed of three tissue layers which are scalp, skull and brain. The reconstruction mesh consists of 50721 nodes associated with 287547 tetrahedral elements, with the average element size 9.3mm$^{3}$. Each node is assigned to one of the three layers. Absorption coefficients assigned to each layer refer to an in vivo study \cite{eggebrecht2012quantitative} at 750nm. 

\begin{figure}[h]
\centering
\includegraphics[width=0.18\textwidth]{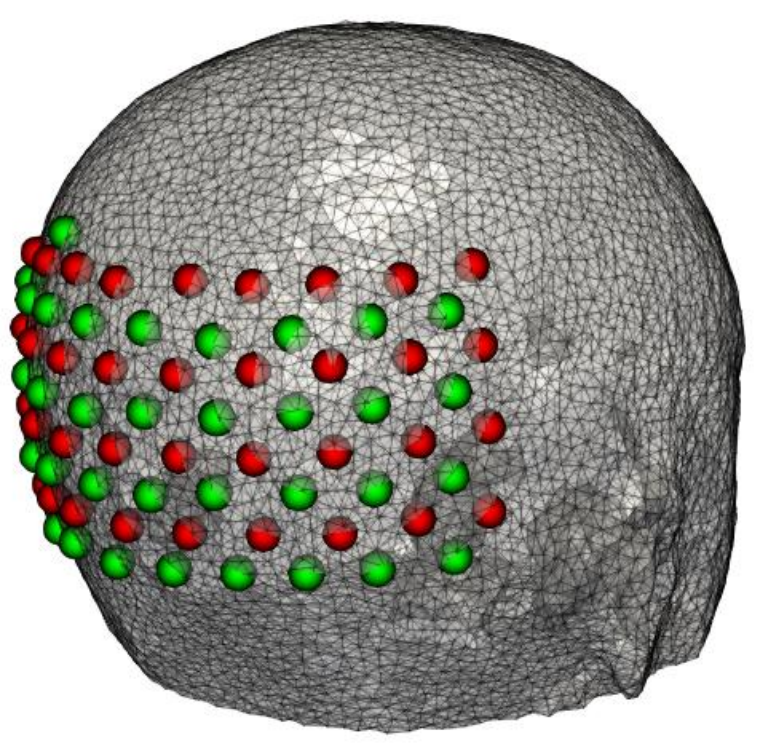}
\caption{Three-dimensional head mesh and distribution of the rectangular imaging array with 36 sources (red dots) and 37 detectors (green dots).}
\label{fig:head1}
\end{figure}

\begin{figure}[h]
\centering
\includegraphics[width=0.5\textwidth]{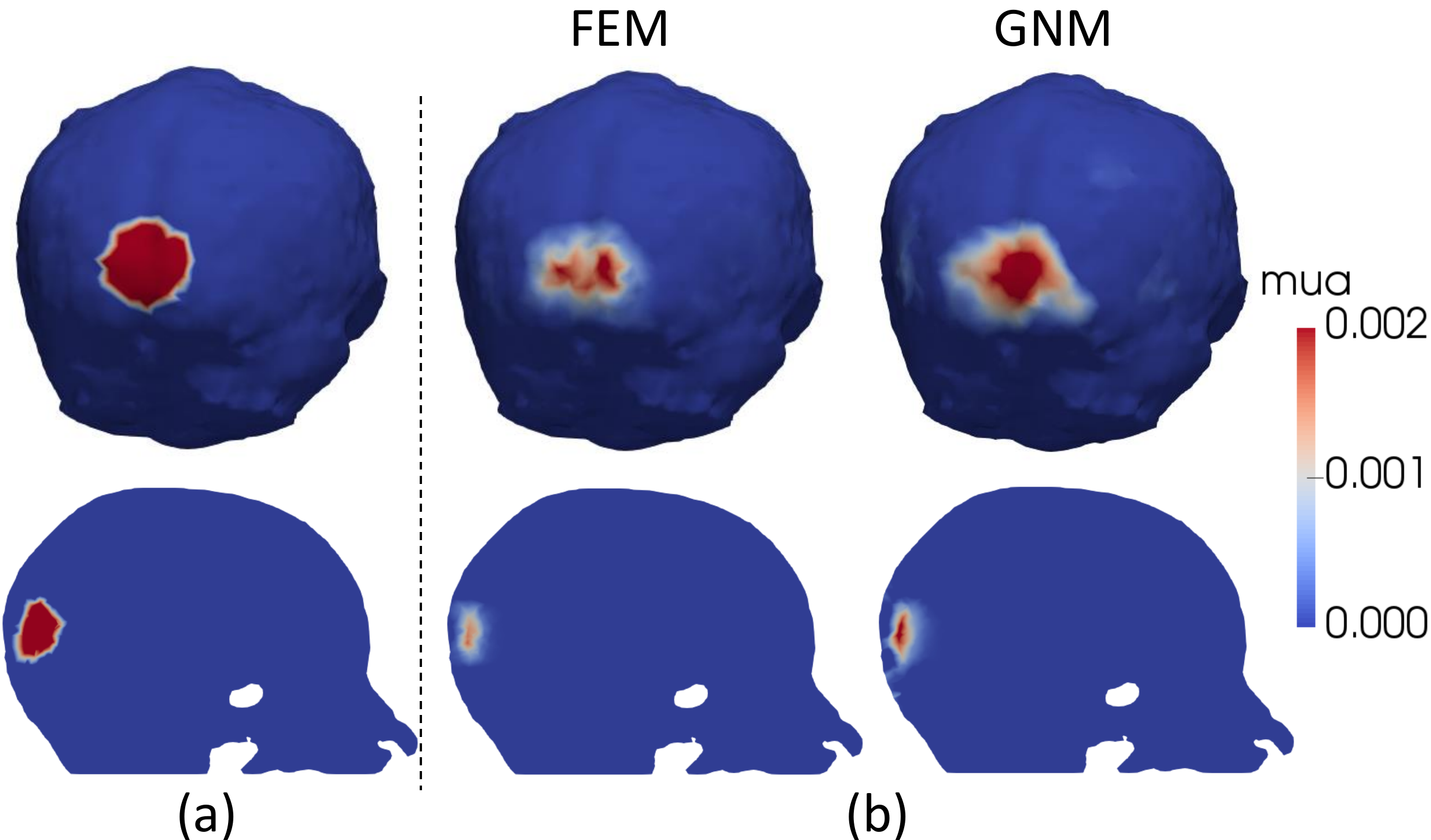}
\caption{(a): Ground truth; (b) Reconstruction with the forward model based on FEM and GNM, respectively.}
\label{fig:head2}
\end{figure}

A large rectangular imaging array with 36 sources and 37 detectors was placed over the back-head area (Figure~{\ref{fig:head1}}), allowing use of multiple sets of overlapping measurements which can improve both the spatial resolution and quantitative accuracy \cite{boas2004diffuse}. The source-detector (SD) separation distances ranges from 1.3 to 4.8cm, leading to 590 overlapping, multi-distance measurements. One anomaly with 15mm radius is simulated in the brain (Figure~{\ref{fig:head2}} (a)). In order to simulate  traumatic brain injury (TBI) cases where the cerebral tissue oxygen saturation ($\rm{StO_2}$) is normally between 50\% and 75\% \cite{ichai2017metabolic} (compared to 80\% in healthy tissue), the absorption coefficient in the anomaly is calculated using Beer's law \cite{dehghani2009near} with 55\% $\rm{StO_2}$. In line with the current in vivo performance of the imaging system, 0.12$\%$, 0.15$\%$, 0.41$\%$ and 1.42$\%$ Gaussian random noise was added to first (13mm), second (30mm), third (40mm) and fourth (48mm) nearest neighbor measurements to provide realistic data \cite{dehghani2009depth}. Reconstructed absorption coefficients of the simulated anomaly using different models are displayed in the second to third column of Figure~{\ref{fig:head2}}. Corresponding 2D cross section is given in the second row. The visualization suggests that GNM can achieve better reconstruction performance with optical property values closer to the ground truth. In addition, the results by both methods are smoothed and the volume sizes of the recovered anomaly are smaller than the ground truth. Evaluation metrics are given in Table~\ref{tb:HEAD_table}. No obvious difference between these two reconstruction models can be observed from the four evaluation metrics. These findings further quantitatively validate the consistency between these two forward models.

\begin{table}[!ht]
\centering 
\caption{\bf Evaluation metrics for $\mu_a$ on the recovered results shown in Figure~{\ref{fig:head2}}.}
\label{tb:HEAD_table}
\begin{tabular}{c|c|c}
\hline
     & \multicolumn{2}{c}{Mean $\pm$ SD}    \\ \hline
     & FEM             & GNM             \\ \hline
AC   & 0.8 $\pm$ 0.0       & 0.9 $\pm$ 0.1      \\ \hline
PSNR & 78.9 $\pm$ 0.0      & 79.0 $\pm$ 0.0      \\ \hline
SSIM & 99.9e-2 $\pm$ 9.1e-7 & 99.9e-2 $\pm$ 3.8e-7 \\ \hline
RMSE & 1.1e-4 $\pm$ 5.1e-7 & 1.1e-4 $\pm$ 1.8e-6 \\ \hline
\end{tabular}
\end{table}

\section{Conclusion}
We have proposed a new formulation of the forward model for DOT that is based on the concepts of differential operators under a nonlocal vector calculus. The discretization of the new forward model is performed using an efficient graph-based numerical method. Our proposed model is shown to be able to accurately model the light propagation in the medium and is quantitatively comparable with both analytical and FEM forward models. Compared with the conventional forward model based on FEM, our proposed model has the following two advantages: 1) according to the experiments in Section 3.1, our proposed model is shown to be more computationally efficient and is up to 64\% faster than the FEM forward model due to the simple graph-based discretization; 2) it allows identical implementation for geometries in different dimensions thanks to the nature of the graph representation.

\section*{Acknowledgment}
The European Union's Horizon 2020 Marie Sklodowska-Curie Innovative Training Networks (ITN-ETN) programme, under grant agreement no 675332, BitMap.

\end{document}